# Why do we find ourselves around a yellow star instead of a red star?

Jacob Haqq-Misra[1,2,*], Ravi Kumar Kopparpu[1,2,3,4], and Eric T. Wolf[5]

[1] Blue Marble Space Institute of Science, 1001 4th Ave, Suite 3201, Seattle, WA
[2] NASA Astrobiology Institute's Virtual Planetary Laboratory, P.O. Box 351580, Seattle, WA
[3] NASA Goddard Space Flight Center, 8800 Greenbelt Road, Mail Stop 699.0 Building 34, Greenbelt, MD
[4] Department of Astronomy, University of Maryland, College Park, MD
[5] Department of Atmospheric and Oceanic Sciences, Laboratory for Atmospheric and Space Physics, University of Colorado Boulder, Boulder, CO

*Email: jacob@bmsis.org

**ABSTRACT**
M-dwarf stars are more abundant than G-dwarf stars, so our position as observers on a planet orbiting a G-dwarf raises questions about the suitability of other stellar types for supporting life. If we consider ourselves as typical, in the anthropic sense that our environment is probably a typical one for conscious observers, then we are led to the conclusion that planets orbiting in the habitable zone of G-dwarf stars should be the best place for conscious life to develop. But such a conclusion neglects the possibility that K-dwarfs or M-dwarfs could provide more numerous sites for life to develop, both now and in the future. In this paper we analyze this problem through Bayesian inference to demonstrate that our occurrence around a G-dwarf might be a slight statistical anomaly, but only the sort of chance event that we expect to occur regularly. Even if M-dwarfs provide more numerous habitable planets today and in the future, we still expect mid G- to early K-dwarfs stars to be the most likely place for observers like ourselves. This suggests that observers with similar cognitive capabilities as us are most likely to be found at the present time and place, rather than in the future or around much smaller stars.

## 1. Introduction

Recent results from the space-based *Kepler* mission as well as ground-based observations have shown that terrestrial planets are about as likely to exist orbiting M-dwarf stars as G-dwarf stars like our sun. The occurrence of rocky planets in the habitable zone around M-dwarf stars is ~20% (Dressing and Charbonneau 2015), similar to the occurrence rate of ~22% for G- and K-dwarf stars (Petigura et al. 2013).[1] M-dwarf stars are more numerous than other stellar types and comprise about 75% of the galactic main sequence stellar population, compared to about 7% for G-dwarf stars. Stars smaller than G-dwarfs also have longer main sequence lifetimes, and the expected lifetime of M-dwarf stars is at least ten times that of our sun. Ground-based surveys have recently found a terrestrial mass planet in the habitable zone of the M-dwarf Proxima Centauri, which is our nearest neighboring star (Anglada-Escudé et al. 2017), as well as three terrestrial mass planets in the habitable zone of the M-dwarf TRAPPIST-1, which is about forty light years away (Gillon et al. 2017). These observations all suggest the possibility that habitable planets are more numerous around M-dwarf stars.

---
[1] Other studies obtained lower estimates for this occurrence rate of 2% (Foreman-Mackey et al. 2014) and 6.4% (Silburt et al. 2015).

The philosophical implications of this bias toward non-G-dwarf stars was expressed by Loeb et al. (2016) as: "The question is then, why do we find ourselves orbiting a star like the Sun now rather than a lower mass star in the future?" The basis for this reasoning is encompassed in the anthropic principle, which requires that the universe must be compatible with our status as conscious observers. Because we find ourselves around a G-dwarf star, even though we predict that K- and M-dwarf stars are more numerous, does this mean we should conclude that we are somehow privileged observers that exist at an unlikely time and place? Such an extraordinary claim would suggest that the emergence of conscious observers on Earth was an extreme statistical fluke or our understanding of habitable environments is significantly flawed.

In this paper we discuss several solutions to this problem that draw upon recent theoretical modeling studies of the habitability of rocky planets around G-, K-, and M-dwarf stars. We demonstrate that our status as observers on a planet orbiting a G-dwarf star is not an improbable statistical anomaly, even if other stellar types are better situated for hosting conscious life.

## 2. Anthropic Reasoning

When considering the factors that led to our emergence as conscious observers on Earth, we tend to expect that we should find ourselves in a typical cosmic environment, that we are more likely average than unusual. This form of reasoning is known as anthropic reasoning, which requires that our observations of the universe are compatible with our existence as conscious living beings. Precursors to anthropic reasoning include the Copernican principle or the principle of mediocrity, which have been used to argue against a geocentric cosmology with Earth's place at the center of the universe in favor of heliocentrism. However, reasoning with the Copernican principle carries the risk of overstating our mediocrity; after all, our existence as conscious, living observers in a universe filled with lifeless planets necessitates that there must be something atypical about our planetary environment. To improve upon Copernican reasoning, Carter (1974) proposed the anthropic principle by arguing that our existence as intelligent, conscious beings on Earth imposes a selection bias when we observe the universe. For example, humans require liquid water to survive, so we should not find it particularly strange that we happen to exist on a wet planet—even if dry planets are more common. Anthropic reasoning leads us in the direction of assuming that Earth is a typical example of an environment suitable for the emergence of conscious observers.

One of the most precise forms of anthropic reasoning is the Self-Sampling Assumption (SSA), which has been thoroughly discussed and critiqued by Bostrom (2002). SSA, as defined by Bostrom (2002), states that, "One should reason as if one were a random sample from the set of all observers in one's reference class." By 'random sample' we mean that we should give equal probability (credence) to finding ourselves as any particular observer in our reference class. By 'set of all observers' we mean all observers that actually exist in the past, present, or future. By 'reference class' we mean the set of entities that count as observers of the universe in the same sense as we do. (We provide an operational definition of reference class in Section 3 below.) SSA allows us to make probabilistic predictions that assume we are average (as if randomly selected) members of our reference class.

With a naive application of SSA, we might conclude that because we find ourselves orbiting a G-dwarf star like the sun, other conscious observers in our reference class (if they exist) must also be more numerous on habitable planets around G-dwarf stars. However, this conclusion is inconsistent with observational evidence suggesting that K- and M-dwarf stars host a greater number of habitable zone planets and may provide more climatologically stable abodes for life against the effects of main sequence brightening (Dressing and Charbonneau 2015; Cuntz and Guinan 2016;

Haqq-Misra et al. 2016). This suggests a more precise expression of the application of SSA to our problem at hand:

> Premise *P1*: Our existence as observers around a G-dwarf star should be considered as a random sample among the set of all observers in G-, K-, and M-dwarf systems.

We will assume *P1* is true in our analysis that follows. We will consider the implications of *P1* on the fact that we exist around a G-dwarf star, rather than a K- or M-dwarf, both at the present era and future of the universe. Because we find ourselves orbiting a G-dwarf star at an early stage of the universe, must we conclude that we are an unlikely statistical fluke? SSA[2] provides us with a tool that, along with astronomical and theoretical constraints, can help to resolve this problem.

## 3. The Reference Class

The issue of defining the reference class of observers is a long-standing problem that we do not purport to solve here; instead, we suggest an operational definition of the reference class as it pertains to our particular scenario. Operational definitions of reference class are commonly used in the search for extraterrestrial intelligence (SETI), where only civilizations that have the ability to utilize radio transmitters have any hope of communicating. In the same vein, we attempt a broader operational definition of reference class for our problem that is based upon cognition, rather than technology.

First, although we talk about the reference class in relation to an individual observer, it is not necessarily the case that all member of the human species will retain the same capabilities of wondering about our position around a G-dwarf instead of an M-dwarf. A person in a coma, for example, cannot wonder about such things due to circumstances, while a young child may lack the full awareness to comprehend such a problem until they become older. Nevertheless, cultures and individuals without strong astronomy education may lack the knowledge to understand the problem immediately, yet such people hold the cognitive capability of understanding the relevant information when taught. The operational reference class of observers for this problem thus seems to require individuals with general unimpaired human cognition along with the ability to comprehend modern scholarly information.

With this constraint in mind, we can better delineate the class of individuals in our operational reference class. The pre-human case is easiest to deal with, as we can imagine a point in the past (such as the emergence of the genus *Homo*) where our ancestors developed the capability to wonder about their place in the universe and comprehend systematic astronomy. Note that it is not required that ancestors in our operational reference class actually performed astronomy or thought about stellar types; rather, it is only relevant that they hold the capability of understanding such knowledge if it were competently taught to them. Prior to this point in history, we assume that hominids earlier than *Homo* were outside of our operational reference class and would be incapable of demonstrating a comparable sort of human-like cognitive ability. We likewise assume that other organisms on Earth are sufficiently different than us in cognitive capabilities so as to place us in a separate operational reference class,[3] as no other organism could seemingly comprehend the implications of modern scholarly information (even if presented in an appropriate form for the particular species).

---

[2] Bostrom (2002) has developed an analog to SSA that specifies one should reason as if one's observer-moment was randomly selected from one's reference class. This Strong Self-Sampling Assumption (SSSA) is an attempt to render ambiguous some of the unpalatable paradoxes of SSA. However, this particular analysis following *P1* contains sufficient indexical information that SSSA effectively reduces to SSA for our scope of interest.

We thus arrive at a definition of operational reference class that applies to this particular problem (even if it lacks more general applicability to other problems in philosophy):

> Premise *P2*: Other observers are in the same operational reference class as us if they possess the cognitive ability to meaningfully comprehend modern scholarly information.

Thus we include humans as well as possible extraterrestrials that SETI could communicate with in our operational reference class, but we exclude rocks, plants, giraffes, and other organisms that lack comparable capabilities as conscious observers. We will assume *P2* is true in the analysis that follows, which will allow us to consider the likelihood that other observers in our operational reference class exist around K- and M-dwarf stars at the present era and future of the universe.

## 4. Analysis with Bayesian Inference

We approach this problem through the framework of Bayesian inference. As an example, consider a fair coin that is tossed three times in a row. Suppose that all three tosses turn up Heads. Can we conclude from this experiment that the coin must be weighted? In fact, we can still maintain our hypothesis that the coin is fair because the chances of getting three Heads in a row is 1/8. Many events with a probability of 1/8 occur every day, and so we should not be concerned about an event like this indicating that our initial assumptions are flawed. However, if we were to flip the same coin 70 times in a row with all 70 turning up Heads, we would readily conclude that the experiment is fixed. This is because the probability of flipping 70 Heads in a row is about $10^{-22}$, which is an exceedingly unlikely event that has probably never happened in the history of the universe. This informal description of Bayesian inference provides a way to assess the probability of a hypothesis in light of new evidence.

We can apply a more formal Bayesian analysis to our problem of inhabited planets in order to estimate the probability of finding ourselves on a planet orbiting a G-dwarf star, rather than an M-dwarf star.[4] We begin by defining the following statements:

> Hypothesis *L* = "A planet is inhabited by conscious life (observers)"
> Evidence *G* = "The planet orbits a G-dwarf star"
> Evidence *K* = "The planet orbits a K-dwarf star"
> Evidence *M* = "The planet orbits a M-dwarf star"

Using these four statements, our goal is to determine the posterior probabilities *P(L|G)* = "The probability that a planet is inhabited by observers given that it orbits a G-dwarf star," *P(L|K)* =

---

[3] Standish (2013) applies this form of reasoning to conclude that 'ants are not conscious' by appealing to SSA: because the population of ants on Earth vastly outnumbers the population of humans, we must conclude that ants are not in our reference class—or else we are left with the conclusion that our birth as humans, rather than ants, is an unlikely statistical fluke.

[4] Our use of Bayesian statistics implicitly assumes that the number of planets inhabited by conscious observers over the history of the universe is a statistically significant quantity. However, if the number of such planets is extremely small (*e.g.*, if Earth is the only inhabited planet across all space and time), then Bayesian methods are not meaningful.

"The probability that a planet is inhabited by observers given that it orbits a K-dwarf star," and $P(L/M)$ = "The probability that a planet is inhabited by observers given that it orbits an M-dwarf star."

In order to determine these posterior probabilities, we must calculate or estimate the likelihoods, which indicate the compatibility of our evidence with our hypothesis. We can express these likelihood quantities as $P(G/L)$ = "The probability that a planet orbits a G-dwarf star given that it is inhabited by observers," $P(K/L)$ = "The probability that a planet orbits a K-dwarf star given that it is inhabited by observers," and $P(M/L)$ = "The probability that a planet orbits an M-dwarf star given that it is inhabited by observers."

We can express the relationship between our likelihood and posterior using Bayes' theorem:

$$P(L|G) = \frac{P(L)P(G|L)}{P(G)}, \tag{1}$$

where $P(L)$ is the probability of conscious observers inhabiting a planet (known as the prior probability) and $P(G)$ is the fraction of G-dwarf stars in the galaxy (known as the model evidence). We can express the fraction of G-dwarf stars as

$$P(G) = \frac{N_G}{N_\star}, \tag{2}$$

where $N_G$ is the total number of G-dwarf stars and $N_\star$ is the total number of stars in the galaxy over all stellar types. We can also express our prior probability for expecting observers elsewhere in the galaxy by invoking the Drake Equation, which we write as

$$P(L) = \frac{N_C}{N_\star}, \tag{3}$$

where $N_C$ is the total number of communicative civilizations in the galaxy for all stellar types. We can then reduce our expression for the G-dwarf posterior probability to

$$P(L|G) = \frac{N_C}{N_G} P(G|L). \tag{4}$$

Likewise, we can write similar expressions for the K-dwarf and M-dwarf posterior probability as:

$$P(L|K) = \frac{N_C}{N_K} P(K|L), \tag{5}$$

$$P(L|M) = \frac{N_C}{N_M} P(M|L). \tag{6}$$

We can now express the ratio of these posteriors, known as the posterior odds, as:

$$\frac{P(L|K)}{P(L|G)} = \frac{N_G}{N_K} \frac{P(K|L)}{P(G|L)}, \tag{7}$$

$$\frac{P(L|M)}{P(L|G)} = \frac{N_G}{N_M} \frac{P(M|L)}{P(G|L)}. \tag{8}$$

Eq. (7) provides us with an expression for the ratio of the probabilities that favor conscious beings observing themselves orbiting a K-dwarf star compared to a G-dwarf star (hereafter known as the 'K-dwarf posterior odds'). Likewise, Eq. (8) expresses the posterior odds that favor observers in an M-dwarf compared to a G-dwarf stellar system (hereafter known as the 'M-dwarf posterior odds'). Note that by using this posterior odds ratio, rather than explicitly using posterior probabilities, the term $N_C$ from the Drake Equation cancels out; our results therefore do not require explicit knowledge of the abundance of intelligent life in the galaxy. We use this formalism in our discussion below that examines several observationally-motivated choices of the likelihood and its implication for the posterior odds of observing ourselves around a G-dwarf, rather than a K-dwarf or M-dwarf, star.

## 5. Spatial Distribution of Observers

We first consider the distribution of observers in space at the present time, assuming that observers are restricted to planets orbiting either G-, K-, or M-dwarf stars. We exclude F-dwarfs and hotter stellar types as candidates for hosting conscious life because we assume that biological evolution requires at least a few billion years of stable main sequence conditions. This allows us to examine three distinct scenarios for planets in the habitable zone: a modest stellar population (G-dwarfs); a large stellar population (K-dwarfs); and a large stellar population where synchronous rotation is expected (M-dwarfs). Actual observations show that each of these categorizations has a wide range of variation, with some late K-dwarfs likely to experience similar problems as M-dwarfs. We neglect these complexities for now, noting that our conclusions for K-dwarfs are particularly applicable to early- to mid-K-dwarfs, while our conclusions for M-dwarfs extend to late K-dwarfs where tidal synchronization is expected.

We consider a planet as being 'habitable' if it is able to sustain standing liquid water on its surface for a long enough duration that could allow for the evolution of life. Our reliance upon calculations of this liquid water habitable zone assumes that life will necessarily require a rocky planet and liquid water to survive. We assume that habitability is a necessary condition for a planet to contain observers, although this need not imply that all habitable planets must be inhabited with observers. We acknowledge that this assumption overlooks additional factors that may contribute to habitability other than the location in the liquid water habitable zone. Various factors contribute to the total number of planets that should be expected around a given planetary system, particularly because planet formation as well as planet migration may show unique behaviors around different stellar types. For example, some predictions such as the packed planetary system hypothesis even suggest that terrestrial planets may be more closely-spaced around smaller stars (Raymond et al. 2009; Kopparapu and Barnes 2010), which might increase the likelihood of inhabited planets around M-dwarfs. TRAPPIST-1 provides observational support for the packed planetary system hypothesis, with seven planets orbiting within 0.063 AU. In addition, we also do not account for exotic biochemistries that might allow life to exist outside the traditional liquid water habitable zone. Likewise, we do not explicitly assess the habitability of exomoons that orbit beyond the limits of the habitable zone, even though tidal heating and planetary illumination could provide energy to such worlds (Heller 2012; Heller and Barnes 2013). Nevertheless, until other indicators for the existence of a habitable planet are forthcoming (for example, evidence from spectroscopic biosignatures), the primary parameter to assess habitability at this time is the location of a planet in the habitable zone of its host star. Our calculations that follow should be considered as illustrative, as the observed number of habitable planets as a function of stellar type will inevitably be updated with future exoplanet characterization missions.

*5.1 We are a statistical fluke at the present time*

Using constraints from habitable zone climate model calculations (Kopparapu et al. 2013, 2014; Leconte et al. 2013; Yang et al. 2013, 2014; Wolf & Toon 2015; Kopparapu et al. 2016; Haqq-Misra et al. 2016), we can calculate the K-dwarf (Eq. 7) and M-dwarf (Eq. 8) posterior odds as a function of the width of the liquid water habitable zone. Beginning with the calculated width of the habitable zone in units of stellar flux, *FHZ*, we can obtain the width of the habitable zone, *HZ*, in terms of physical radial distance as $HZ_i = \sqrt{l_i/FHZ_i}$, where *l* is stellar luminosity and $i \in \{G, K, M\}$ denotes stellar type. Using this relationship, we write the likelihood function in a general form as:

$$P(i|L) = \frac{N_i HZ_i}{N_G HZ_G + N_K HZ_K + N_M HZ_M}, \qquad (9)$$

where $HZ_G$, $HZ_K$, and $HZ_M$ are the respective calculated widths of the habitable zone (in units of radial distance). For the G-dwarf likelihood (*i* = *G*), Eq. (9) implies that the probability that a planet orbits a G-dwarf star, given the planet is inhabited, depends upon the fractional width of the G-dwarf habitable zone (i.e, how much habitable 'real estate' is available around a G-dwarf?) and the relative fraction of the G-dwarf stars in our galaxy (i.e, how common are G-dwarf stars?). Similarly, Eq. (9) expresses the K-dwarf (*i* = *K*) and M-dwarf (*i* = *M*) likelihood functions. We limit our consideration in this analysis to three stellar types, which requires that *P(G|L) + P(K|L) + P(M|L) = 1*.

We can use model calculations for the inner edge (Kopparapu et al. 2013, 2014; Yang et al. 2013, 2014; Leconte et al. 2013; Wolf & Toon 2015; Kopparapu et al. 2016) and outer edge (Kopparapu et al. 2013, 2014; Haqq-Misra et al. 2016) of the habitable zone to find $HZ_G$, $HZ_K$, and $HZ_M$. We first note that one-dimensional climate models (Kopparapu et al. 2013, 2014) find the inner edge of the habitable zone for a G-dwarf at $FHZ_{G,inner} \approx 1.1 S_0$ and the outer edge at $FHZ_{G,outer} \approx 0.35 S_0$ (where $S_0$ is the present-day value of solar flux). For K-dwarfs, results from one-dimensional climate models as well as more complex general circulation models (Kopparapu et al. 2016; Haqq-Misra et al. 2016) show that the inner edge is at $FHZ_{K,inner} \approx 1 S_0$ and the outer edge is at $FHZ_{K,outer} \approx 0.26 S_0$. For M-dwarfs, we focus on results from general circulation models (Yang et al. 2013, 2014; Kopparapu et al. 2016) because such models are able to quantify the three-dimensional aspects of atmospheric circulation on a synchronously rotating planet. These calculations show that planets in synchronous rotation around an M-dwarf have an inner edge at $FHZ_{M,inner} \approx 1.5 S_0$ and an outer edge at $FHZ_{M,outer} \approx 0.25 S_0$. If we assume that $l_K = 0.1 l_G$ and $l_M = 0.01 l_G$ as nominal values, then we can calculate the respective habitable zone widths for G-, K-, and M-dwarfs in units of radial distance as: $HZ_G = \sqrt{1/0.35} - \sqrt{1/1.1} = 0.74$ AU, $HZ_K = \sqrt{0.1/0.26} - \sqrt{0.1/1.0} = 0.30$ AU, and $HZ_M = \sqrt{0.01/0.35} - \sqrt{0.01/1.5} = 0.12$ AU. Note that even though the habitable zone is much larger for K- and M-dwarfs when measured in stellar flux (*FHZ_{outer} – FHZ_{inner}*) compared to G-dwarfs, the actual habitable zone widths are comparably smaller for K- and M-dwarfs because of their lower luminosity.

We know from observations that $N_G/N_\star = 0.07$, $N_K/N_\star = 0.12$, and $N_M/N_\star = 0.77$. We then obtain values for our three likelihoods: *P(G|L)* ≈ 0.29, *P(K|L)* ≈ 0.20, and *P(M|L)* ≈ 0.51. This reduces our expressions for posterior odds in Eqs. (7) and (8) to:

$$\frac{P(L|K)}{P(L|G)} \approx 0.40, \tag{10}$$

$$\frac{P(L|M)}{P(L|G)} \approx 0.16. \tag{11}$$

These results do not require us to discard our assumed likelihoods, nor do we need to reject anthropic reasoning. We are much less likely to favor our existence around an M-dwarf star, so the fact that we instead exist around a G-dwarf star remains consistent with our calculated posterior odds. We therefore arrive at our first conclusion, with the assumption that G-, K-, and M-dwarf stars are equally probable hosts for conscious observers at the present time:

> Conclusion $C1$: Our existence around a G-dwarf star at the present era is expected, with about a ninety percent lower likelihood of instead existing around an M-dwarf star today.

Although M-dwarfs are much more numerous than G-dwarfs, the larger physical space of the habitable zone ($HZ_G > HZ_M$) makes G-dwarfs slightly more likely for hosting conscious observers. For observers across space at the present time, $C1$ suggests that we can reason that our existence around a G-dwarf star is to be expected, even if K- and M-dwarfs provide numerous habitable planets.

*5.2 Observers are less numerous on planets orbiting M-dwarf stars at the present time*
One possibility is that observers are less numerous on planets orbiting M-dwarf stars. A variety of factors such as synchronous rotation, stellar flaring activity, and water loss during pre-main sequence evolution could exacerbate or preclude life on such planets. If we accept this class of arguments as valid, then this implies that the M-dwarf posterior odds are less than the G-dwarf and K-dwarf posterior odds.

Historically, the problem of synchronous rotation for planets within the habitable zone of M-dwarfs has suggested that such planets are uninhabitable. Dole (1964) first argued that the atmosphere of a synchronously rotating planet is prone to freeze out on the permanent night side of the planet, which would render the planet an airless and uninhabitable ball of ice. This line of reasoning concludes that M-dwarf planets are likely to be uninhabitable, even if they are within the habitable zone, so that $P(M|L) \approx 0$. This implies that the best candidates for life are planets around G- and K-dwarf stars where synchronous rotation is not a problem within the habitable zone. However, Haberle et al. (1996) used a simplified climate model to argue that energy transport from the day to night side of a synchronously rotating planet is sufficient to keep even thin atmospheres from collapsing. This result was substantiated with more sophisticated calculations using general circulation models by Joshi et al. (1997) and others (Joshi 2003; Merlis and Schneider 2010; Edson et al. 2011, 2012; Carone et al. 2014; Yang et al. 2013, 2014; Kopparapu et al. 2016), which all indicate that a wide range of synchronously rotating atmospheres are able to withstand collapse. Synchronous rotation itself seems to pose less of a problem for habitability than originally proposed by Dole (1964) and others.

More recent investigations have highlighted other problems that could preclude the development of life around M-dwarf planets. Most notable is that planets forming around M-dwarf stars are likely to find themselves in a runaway greenhouse state during the hot and extended pre-main sequence phase of the host star (Ramirez and Kaltenegger 2014; Luger and Barnes 2015; Tian and Ida 2015). Unless such planets are able to regain water through delivery by impactors at a later stage, this problem of pre-main sequence water loss could prevent the development of life on many M-dwarf planets. Other problems include flares from the enhanced stellar activity of M-dwarfs (Segura et al. 2010) or the possible deficiency in volatiles for such planets during formation (Lissauer 2007). Even though M-dwarfs are more numerous, complications such as these could lead us to conclude that other factors favor the emergence of conscious observers around G- and K-dwarf stars.

However, several studies offer some respite to these problems. Yang et al. (2013) find that slow and synchronously rotating planets develop thick substellar cloud decks which strongly cool the planet and may in part, help mitigate pre-main sequence water-loss. Furthermore, Tian and Ida (2015) argue that M-dwarf planets may form in two primary modes, either as dry desert planets or as waterworlds with a much greater oceanic volume than Earth. Desert planets can remain habitable under a wider range of stellar fluxes because water vapor and sea-ice albedo feedbacks are nonexistent (Abe et al. 2011; Leconte et al. 2013), although such planets may be less desirable candidates for hosting life. Deep water-world planets may allow extended periods of water-loss without fully desiccating the planet, still leaving an ocean behind after the pre-main sequence phase, and pose no problems to the development of advanced life. Theoretical photochemical modeling studies indicate that even strongly flaring M-dwarf stars, like AD Leo, would not present a direct hazard for life on planet the surface, given an Earth-like atmosphere with $O_2$ and $O_3$ that provides an ultraviolet shield (Segura et al. 2010). These arguments and others (Tarter et al. 2007) all suggest that at least some planets orbiting M-dwarfs could develop advanced forms of life (in which case the conclusion *C1* is still justified).

Appealing to physical limitations on habitability does solve the problem of why we do not orbit an M-dwarf star. If we accept these arguments as valid, then we can assume $P(M/L) \approx 0$, which considers the limiting case in Eq. (9) where M-dwarf planets are completely uninhabitable. This assumption yields three new values for our likelihoods: $P(G/L) \approx 0.59$, $P(K/L) \approx 0.41$, and $P(M/L) \approx 0$. We can then calculate new posterior odds with our pessimistic assumptions about M-dwarf habitability:

$$\frac{P(L|K)}{P(L|G)} \approx 0.41, \qquad (12)$$

$$\frac{P(L|M)}{P(L|G)} \approx 0. \qquad (13)$$

These results not only resolve the M-dwarf observer problem, but they also adjust our posterior odds for K-dwarfs. Most of the physical constraints on habitability for planets orbiting M-dwarf stars (such as synchronous rotation and stellar activity) do not apply to the early- to mid-K-dwarf systems that we also consider. This leads to our second conclusion, with the assumption that M-dwarf stars are much less probable hosts for conscious observers than G- and K-dwarfs at the present time:

> Conclusion *C2*: If M-dwarfs are less hospitable hosts for habitable planets, then our existence around a G-dwarf star at the present era is to be expected, with nearly half the likelihood of instead existing around a K-dwarf star today.

Thus, if we assume that M-dwarf planets are problematic for life, then we should further consider it typical that we find ourselves orbiting a G-dwarf star, with a slightly less likelihood of instead orbiting a K-dwarf star.

## 6. Temporal Distribution of Observers

We next consider the distribution of observers in space and across all time, still assuming that observers are restricted to planets orbiting in the liquid water habitable zone either G-dwarf, K-dwarf, or M-dwarf stars. Conscious life on Earth took about 4.5 billion years to develop after the planet formed, but with a sample size of one we cannot predict how typical this timescale may be for other stellar environments. The longer main sequence lifetime of K- and M-dwarfs provides additional time for the requisite biological evolution and the development of conscious observers. We must therefore extend our framework to consider the likelihood that we should exist on a G-dwarf star today, rather than a K- or M-dwarf star in the distant future.

### 6.1 We are a statistical fluke across all time

The first step in considering the distribution of observers across time is to modify our argument in subsection 5.1 to account for the expected lifetime of each stellar type. We assume that habitability is limited to the main sequence lifetime of a star, with a typical lifetime of $t_G = 10^{10}$ yr for G-dwarf stars, $t_K = 10^{11}$ yr for K-dwarf stars, and up to $t_M = 10^{13}$ yr for small M-dwarf stars. We consider the distribution of habitable planets across the entire history of the universe from time $t = 0$ to $t = t_M$, which is the timeframe when conscious observers could arise. We can write the fractions of the age of the universe that each stellar type remains in its main sequence phase as $f_G = t_G/t_M = 10^{-3}$, $f_K = t_K/t_M = 10^{-2}$, and $f_M = t_M/t_M = 1$. We can then write a new expression for our likelihood function that takes into account this temporal effect:

$$P(i|L) = \frac{f_i N_i HZ_i}{f_G N_G HZ_G + f_K N_K HZ_K + f_M N_M HZ_M}, \tag{14}$$

This gives three new values for our likelihoods: $P(G|L) \approx 5.6 \times 10^{-4}$, $P(K|L) \approx 3.9 \times 10^{-3}$, and $P(M|L) \approx 0.99$. We can now express the K-dwarf and M-dwarf posterior odds for habitable planets across time as:

$$\frac{P(L|K)}{P(L|G)} \approx 4.0, \tag{15}$$

$$\frac{P(L|M)}{P(L|G)} \approx 160. \tag{16}$$

This leads to our third conclusion, with the assumption that G-, K-, and M-dwarf stars are equally probable hosts for conscious observers across all time:

> Conclusion *C3*: Our existence around a G-dwarf star is unlikely, with over a hundred times greater likelihood of existing around an M-dwarf star in the future.

This does not necessarily mean that anthropic reasoning has failed, but only that we have found ourselves in a less than average, although not wholly improbably, stellar environment. The M-dwarf posterior odds tell us that we are over 100 times more likely to find ourselves around an M-dwarf star than a G-dwarf star. This is a less optimistic number than our present-day estimate with Eq. (11), but does this mean that we must reject our assumed likelihoods? Is a 1 in 100 chance reasonable within anthropic considerations, or do must we reject anthropic reasoning here? In fact, plenty of events occur all the time with a chance of 1 in 100 or more. Approximately one percent of people are ambidextrous, while nearly the same fraction are born with red hair. Scaling further up, for sake of argument, could we accept even a 1 in 10,000 chance to account for our existence around a G-dwarf? The odds are about 1 in 10,000 that a person has been bitten by a Malaria-carrying mosquito or can sing with perfect pitch. About 1 in 10,000 people are involved in a flood, a landslide, or a grizzly bear attack. All of these events are realistic and occur on a regular basis, even if not a daily basis. Thus, we need not necessarily reject anthropic reasoning or our likelihoods because it also remains plausible that conscious life on Earth is a one-chance-in-a-hundred fluke.

### *6.2 Observers are unlikely on planets orbiting M-dwarf stars across all time*

To complement our discussion from subsection 5.2, we consider the possibility that planets orbiting M-dwarf stars may be unsuitable environments for life, either now or at any point in the future. If M-dwarf planets are completely inhospitable to life, then $P(M|L) \approx 0$ and the M-dwarf posterior odds are zero. As in subsection 5.2, this assumption yields three new values for our likelihoods: $P(G|L) \approx 0.13$, $P(K|L) \approx 0.87$, and $P(M|L) \approx 0$. We can then calculate new posterior odds with our pessimistic assumptions about M-dwarf habitability over the history of the universe:

$$\frac{P(L|K)}{P(L|G)} \approx 3.9, \tag{17}$$

$$\frac{P(L|M)}{P(L|G)} \approx 0. \tag{18}$$

We thus arrive at our fourth conclusion, with the assumption that M-dwarf stars are much less probable hosts for conscious observers than G- and K-dwarfs across all time:

> Conclusion *C4*: If M-dwarfs are less hospitable hosts for habitable planets, then our existence around a G-dwarf star is a slight statistical fluke, with about a four times greater likelihood of existing around a K-dwarf star in the future.

We need not be concerned about a one in four chance for our existence around a G-dwarf compared to a K-dwarf. Nearly one in four people in the world follow Islam, while about one in five live in China; such characteristics are commonplace, even if they do not describe the majority of people. It is tempting to conclude based on physical reasons, as well as the posterior odds from Eq. (16), that conscious observers should not exist on planets orbiting M-dwarf stars. While we can appeal to small deviations from the mean to explain our position around a G-dwarf star instead of a K-dwarf star, we require even more extraordinary circumstances to explain why we do not exist around an M-dwarf star in the future. Rather than accept the improbable outcome of Eq. (16), perhaps the combination of anthropic reasoning and climate research suggests that assumptions like those in Eq. (18) are correct. If M-dwarfs planets really do experience a range of problems that limits their potential for developing life, then we can reason that our place around a G-dwarf today is a common environment for observers to find themselves.

## 7. Discussion

If this analysis had turned up a much larger number, such as a 1 in $10^{22}$ chance of us finding ourselves around a G-dwarf star, then of course we would have to reject either our likelihoods or anthropic reasoning, because we essentially never expect to observe events with a 1 in $10^{22}$ chance. But this is not what we have found, so we need not be concerned about finding ourselves in a less-than-typical cosmic environment. So, to return to the question at the title of this paper: We find ourselves orbiting a G-dwarf star because we are a little bit of a statistical anomaly, but only the sort of anomaly that we experience on a regular basis.

This still leaves open the degree of atypicality we are claiming for ourselves. Understanding the habitability of M-dwarf planets is critical to resolving this ambiguity: if M-dwarf planets are poor candidates for life, then the distant future ten trillion years from now may be a worse time for the emergence of conscious observers in our reference class than the present era. Theoreticians have identified several possible problems that could preclude the development of life on M-dwarf planets, although others have suggested plausible resolutions that could still maintain habitable conditions. One interpretation of this state of affairs is that climate models contain too many Earth-like assumptions, and M-dwarf habitability problems tend to disappear when we develop more robust tools for simulating a wider range of planetary atmospheres. Another interpretation is that our knowledge of planetary systems is functional enough that we can claim with confidence that M-dwarf systems are poorer environments for life than G-dwarf systems. Observational resolution of this dilemma will be difficult and may only be possibly when spectroscopic measurements of the atmospheres of M-dwarf planets become possible with the next generation of space telescopes.

Further observations pending, the best we can do is appeal to SSA and claim that our present status as conscious observers may as well be randomly selected from the set of total observers in our reference class (across the entire past, present, and future of the universe). We find ourselves around a G-dwarf star at an early stage in the universe, which according to *C3* is disfavored by a factor of ~100 to our existence around an M-dwarf star. Although claiming our existence to be a 1 in 100 statistical fluke remains within the realm of possibility, is such a conclusion philosophically satisfying? Or should we take this as an indication that our physical intuitions about M-dwarf habitability may also be correct? Perhaps our combined expectations from planetary habitability theory and anthropic reasoning with SSA both point toward the conclusion *C4* that M-dwarf planets really are more difficult places for life.

Another perspective on this problem derives from the doomsday argument (Carter and McCrea 1983; Gott 1993; Bostrom 2002), which considers the longevity of our current reference class as observers. Succinctly, the doomsday argument supposes that if humans are born in a random order, we should assume from SSA to be sequentially somewhere near the middle of the population distribution. When coupled with our rapid exponential growth in population, this suggests that we will reach the predicted total number of humans fairly soon in the future. Thus, so the argument goes, doomsday (of one sort or another) becomes more probable when one considers their birth rank as typical (*i.e.*, randomly selected as with SSA). If the doomsday argument is correct, then human civilization as we know it is much closer to its end than its beginning.

The doomsday argument sounds objectionable to many at first and has generated broad discussion both in favor and against its conclusions (*e.g.*, Korb and Oliver 1998; Bostrom 1999; 2002; Olum 2002; Bostrom and Ćirković 2003; Monton 2003). We do not necessarily endorse the strong conclusions of the doomsday argument here; instead, we note that the prospect of so-called doomsday need not correlate to a global catastrophe or existential threat. In general, any significant transformation that places post-humans in a different reference class as humans today would qualify as satisfying the criterion for doomsday. The concept of a post-human, or transhuman, can refer to any biological or technological development in the future that results in humans now evolving into something notably different than (and probably reproductively incompatible with) humans today. For example, we do not include animals such as cats or owls in our operational reference class because such organisms hold a limited set of cognitive capabilities to qualify as observers on the same terms as us. Likewise, we can imagine a technological or biological development that places post-humans in a new operational reference class. One form of such transformation is the suggestion by Kurzweil (2005) and others that humans will soon be able to transfer our biology-based intelligence into robot brains that live much longer. Another possibility suggested by Grinspoon (2009) and others is that the universe is populated by 'immortal' long-lived civilizations that have transcended their problems of sustainable development. Perhaps more likely, a shift in our operational reference class will be due to long-term genetic changes that result when any biological creature evolves over time. Just as we can envision a point in the past when distant pre-human ancestors would not be considered part of our operational reference class, so too can we imagine a threshold where distant post-human descendants are also no longer considered part of our operational reference class. At some point in the future, humans will be drastically different than we are now.

Haussler (2016) raises the possibility that our unlikely position at this early phase of the universe provides confidence that we are entering an era of 'interstellar convergence,' where emergent interstellar communication networks between extraterrestrial civilizations in the galaxy will soon grow powerful enough that they become a permanent feature of future civilizations. Haussler (2016) argues that we are more likely to find ourselves wondering about the existence of extraterrestrial life if we exist at a stage in the universe's history when interstellar communication between civilizations is just beginning. By contrast, if such interstellar convergence never occurs, or if it occurs in the distant future, then we would expect to find ourselves at a more typical era in the universe's history, much later in the future (and probably around an M-dwarf star). Thus, Haussler (2016) claims that the success of SETI and the establishment of an era of interstellar convergence will propel humanity into our next evolutionary state—and, necessarily, into a different operational reference class. According to this logic, we find ourselves around a G-dwarf star at an early phase in the universe because this is the most likely place and time to expect observers in our operational reference class.

Although compelling as a motivation for SETI, the argument of Haussler (2016) cannot distinguish between events that positively transform humanity (such as entering an era of interstellar convergence) and events that negatively transform humanity (such as a literal doomsday that ends with extinction). Whether or not interstellar convergence is likely, it remains possible that the operational reference class of observers itself is a function of time in the universe. If observers like us (*i.e.*, in our operational reference class) do not exist in the future, then we are left with the present-day spatial conclusions *C1* and *C2*, which suggest that our position around a G-dwarf is reasonably typical.

Whether by biological evolution or technological enhancement, it seems inescapable to conclude that post-humans will be at least as cognitively capable as humans, and probably much more so. We therefore cannot exclude post-humans from our operational reference class on the condition of less-than-human cognitive abilities. Instead, we also suppose that post-humans will have access to a greater wealth of information than available today, which could render obsolete some of our current modes of thinking. For example, the scenario envisioned by Haussler (2016) suggests that the advent of widespread interstellar communication could lead to a transformation of our operational reference class; observers born into such a universe would have no need to wonder about their relationship to others because such information would be known. Another possibility could revolve around the temporal perception of a conscious being (DeVito 2011). Humans have lifespans of ~100 years with daily activities on the order of hours and thoughts that occupy seconds to minutes. Post-humans that evolve to operate on different timescales may fall into different modes of conscious thought than our current operational reference class. Other imaginative and mundane possibilities can be explored, but the point remains that our present operational reference class may be short-lived even if humans evolve into a post-human species.

The idea that the operational reference class of observers may depend upon the relevant timescale for the species in question deserves additional consideration. G-dwarf stars are shorter lived than M-dwarfs, but G-dwarfs have a greater luminosity during the main sequence phase. Perhaps the evolution of life on planetary systems develops on a timescale proportional to the lifetime of the star itself, so that M-dwarfs today should not expect to develop any form of advanced life until much later in the future. This suggests the possibility that the operational reference class of observers depends upon stellar type, perhaps as a result of life developing to prefer different timescales of operation on G-dwarfs compared to M-dwarfs. If this line of reasoning is valid at all, then we should not expect to find signs of conscious observers on M-dwarf planets today, while G-dwarf and early K-dwarf planets provide better places to look. This line of reasoning leads to our fifth conclusion:

> Conclusion $C5$: If observers in our operational reference class do not exist in the future, then our existence around a G-dwarf star at the present era is to be expected.

We therefore find ourselves orbiting a G-dwarf star in an early phase of the universe because this is precisely where we expect observers in our operational reference class should exist.

## 6. Conclusion

Our place around a G-dwarf star, instead of a K- or M-dwarf, is a typical occurrence that requires no special circumstances to explain. Even if we accept that M-dwarf planets provide more numerous habitable sites across the history of the universe, our existence around a G-dwarf star would only be the sort of statistical fluke that occurs on a routine basis. Even so, our application of anthropic reasoning provides corroborating evidence to theoretical expectations that suggest M-dwarf habitability may be limited. If we allow for at least some reduction in the suitability of M-dwarf planets for life, then we arrive at an even stronger conclusion that our position on a planet around a G-dwarf is reasonably typical.

Our analysis also depends upon the idea that other observers in our operational reference class are likely to exist around other stellar types. But this assumption could be flawed, particularly if the operational reference class of observers is itself a function of stellar luminosity, and therefore stellar type. Post-humans, too, may transition toward a new operational reference class—perhaps through a breakthrough in technology, perhaps through initiation of contact with extraterrestrials, or perhaps (although hopefully not) through the extinction of humanity. If we accept that our current operational reference class status is finite, and if we further accept that the concept of reference class itself may be linked to stellar type, then we are left with the conclusion that the present time and place is optimal for expecting observers like us.

**Acknowledgments**

The authors thank Gerry Harp, Eddie Schweiterman, and Jason Wright for thoughtful comments that greatly improved the manuscript. J.H.-M., R.K.K., and E.T.W. gratefully acknowledge funding from the NASA Habitable Worlds program under award NNX16AB61G. J.H. received additional funding from the NASA Habitable Worlds program under award NNX15AQ82G. R.K.K. also received funding from NASA Astrobiology Institute's Virtual Planetary Laboratory lead team, supported by NASA under cooperative agreement NNH05ZDA001C. Any opinions, findings, and conclusions or recommendations expressed in this material are those of the authors and do not necessarily reflect the views of NASA.